\documentclass[pra,twocolumn]{revtex4-1}
\usepackage{amsmath, amssymb, graphicx}
\usepackage{hyperref}
\usepackage{soul,color}
\usepackage{siunitx}
\usepackage[utf8]{inputenc}

\newcommand{\modelabel}{\mathfrak{m}}
\newcommand{\omDye}{\epsilon}
\renewcommand{\vec}[1]{\mathbf{#1}}
\newcommand{\dsolid}{\iint d\Omega}
\newcommand{\ntot}{n_{\text{tot}}}

\begin{document}

\title{Polarization dynamics in a photon Bose-Einstein Condensate}

\author{Ryan I. Moodie}
\affiliation{SUPA, School of Physics and Astronomy, University of St Andrews, St Andrews, KY16 9SS, United Kingdom}
\author{Peter Kirton}
\affiliation{SUPA, School of Physics and Astronomy, University of St Andrews, St Andrews, KY16 9SS, United Kingdom}
\author{Jonathan Keeling}
\affiliation{SUPA, School of Physics and Astronomy, University of St Andrews, St Andrews, KY16 9SS, United Kingdom}
\date{\today}

\begin{abstract}
  It has previously been shown that a dye-filled microcavity can produce a
  Bose-Einstein condensate of photons. Thermalization of photons is possible via repeated absorption and re-emission by the dye molecules.  In
  this paper, we theoretically explore the behavior of the polarization of light in
  this system.  We find that in contrast to the near complete thermalization
  between different spatial modes of light, thermalization of polarization
  states is expected to generally be incomplete.  We show that the polarization degree
  changes significantly from below to above threshold, and explain
  the dependence of polarization on all relevant material parameters.
\end{abstract}

\maketitle

\section{Introduction}
\label{sec:introduction}

By trapping photons in a high quality multimode resonator, and allowing
them to interact with emitters such as dye molecules, it is possible to
form a thermalized gas of photons, and at high enough densities, a
Bose-Einstein condensate~\cite{Klaers2010b,Klaers2010c}.  The crucial
feature in these experiments is the complex spectrum of typical dye
molecules: there are broad absorption and emission spectra, and these
spectra are related to each other by a Boltzmann factor (a feature known as
the Kennard-Stepanov relation~\cite{Kennard1918,Kennard1926,Stepanov1957}).
This behavior arises because the internal rovibrational state of the
molecules rapidly reaches thermal equilibrium due to collisions between dye
molecules and the solvent.  In turn, this Boltzmann factor between emission
and absorption leads to a Bose-Einstein distribution of the photon energy, as long as they can be absorbed and re-emitted many times before
escaping the cavity.

The observation of Bose-Einstein condensation of photons has lead to many
subsequent theoretical discussions and experimental
extensions~\cite{Klaers2012a,Sob'yanin2012,Snoke2012,Kirton2013b,Kruchkov2014,VanderWurff2014,DeLeeuw2014,Chiocchetta2014,Nyman2014,Sela2014,Kirton2015,DeLeeuw2014a,Schmitt2014b,Schmitt2014,Chiocchetta2015,Marelic2015,
  keeling16, marelic16:njp,marelic16,klaers2016bose, hesten2017,nyman17small}, based in part on the
simplicity of the material system.  Here we focus on an aspect that has
only recently been cursorily studied~\cite{Jagers2016,Nyman:PolPrivComm},
the dynamics of polarization.

Polarization dynamics has of course been extensively studied for other
Bose-Einstein condensates, for a variety of reasons.  In ultracold atoms, a
number of questions have been
studied~\cite{stenger1998,sadler2006} such as the dynamics
following a sudden quench, leading to the formation of domains, and
non-trivial coarsening dynamics --- for a review of these topics, 
see e.g.~\citet{stamper-kurn13}.
In Helium, there can be a complex spinor
structure arising from spin-orbit coupling, and this in turn can lead to
complex topological defects~\cite{Salomaa1987b,Leggett:QL}.  Most directly
relevant to the photon BEC system is the study of polarization dynamics in
polariton condensates, which has been reviewed by~\citet{shelykh10}.

For both photon and polariton condensates, the polarization state of a
condensate can serve as a clear test of the relative roles of energetics vs
the balance of pumping and decay.  Were the system to fully reach
equilibrium, the polarization state would be entirely determined by
any energetic splitting between polarized states.  
Experiments on polaritons~\cite{larionov10} have however seen
cases where the higher energy of two Zeeman split states becomes
macroscopically occupied.  Such observations are often clearer than
the equivalent behavior in terms of competition between different spatial
modes: the polarization state can be described by a small number of
parameters, and so fully quantified.  However, in polariton systems this
is complicated by an intrinsic splitting between different linear
polarization states which arises from the quantum well
structure~\cite{klopotowski06,kasprzak07}.  Nonetheless, theoretical
models have been developed~\cite{laussy06,solnyshkov09}
that compare well to experiments.

In this paper, we develop a model for the polarization dynamics of a dye-filled
microcavity.  We show that in contrast to the spatial dynamics,
which is expected to fully thermalize in the limit of a perfect cavity~\cite{keeling16}, the polarization state need not do
so.  This means that the output polarization depends on the polarization
of the pump light.   This dependence varies significantly from below
to above threshold.  Below threshold, rotational diffusion of molecules
destroys polarization, while above threshold, stimulated emission overcomes
this, leading to a highly polarized state.  
The model we develop  accounts for specific dynamics of the photon 
condensate, namely the repeated absorption and re-emission of light by
dye molecules.  This requires developing coupled equations for the
polarization state of the light in the cavity, as well as for the state
of the dye molecules.  Such a model is thus quite different from
the typical situations encounted in e.g. atomic condensates~\cite{stenger1998,sadler2006}.

Our paper is organized as follows.  Section~\ref{sec:model} introduces the
model we consider, and provides a discussion of the parameters which appear.
We then use this model in Section~\ref{sec:results} to explore
the dependence on various parameters.  First, to orient the reader, we
illustrate typical states found below and above threshold (i.e.\ showing the
occupations of all spatial and polarization modes).  We then proceed to
extract a net polarization of the output, and explore the dependence of this
characteristic quantity upon all relevant parameters.  Finally, in
Section~\ref{sec:conclusions} we summarize our results.

\section{Model}
\label{sec:model}

\begin{figure}[b]
  \centering
  \includegraphics[width=3.2in]{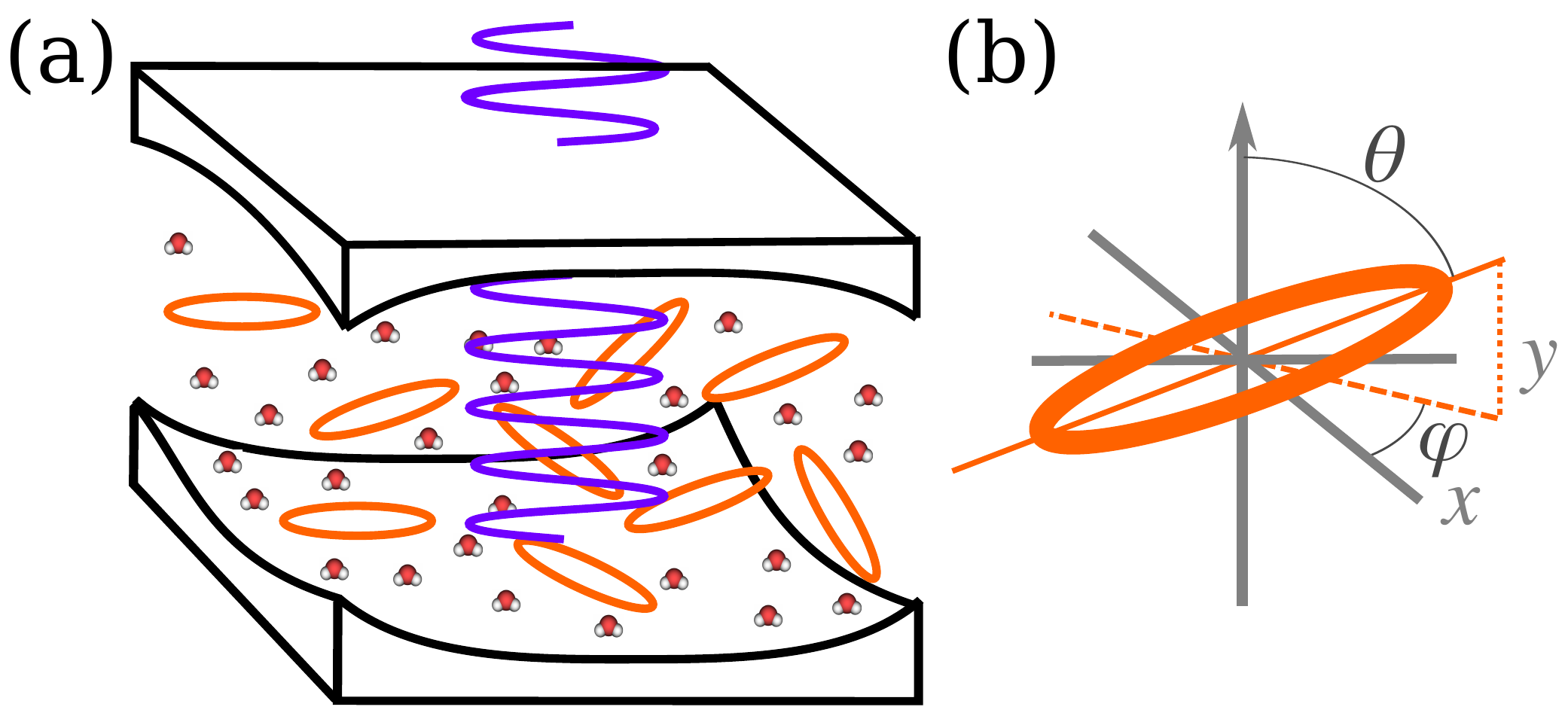}
  \caption{(a) Cartoon illustrating the physical system we consider, and (b)
    the angles defining the orientation of a molecule.  Orange ellipses
    represent dye molecules, indicating the principal axes of the
    electric dipole, sitting in a background of solvent molecules.  
    A fully $y$ polarized pump is shown in (a).}
  \label{fig:cartoon}
\end{figure}


\subsection{Underpinning Hamiltonian}
\label{sec:underp-hamilt}

Our basic starting point is a modified form of the multimode Jaynes-Cummings model
\begin{multline}
  \label{eqn:H0}
  \hat{H}=\sum_{\modelabel}\omega_\modelabel\hat{a}^\dagger_\modelabel \hat{a}^{}_\modelabel 
  +
  \sum_i {\omDye} \sigma^+_i \sigma^-_i
  +
  g\sum_{\modelabel,i}\left(\hat{a}^{}_\modelabel\hat{\sigma}_i^++\hat{a}^\dagger_\modelabel\hat{\sigma}_i^-\right)
  \\
  + H_{\text{rovibrational}}
\end{multline}
using units such that $\hbar=1$.  As discussed
previously~\cite{Kirton2013b,Kirton2015}, the thermalization of photons by dye
fluorescence can be modeled by considering how the electronic transitions of a
molecule are dressed by coupling to internal vibrational degrees of freedom of
the molecule.  After a polaron transform, these degrees of freedom appear as
dressing the electronic transitions, by the replacement $ \sigma^+_i \to
\sigma^+_i \hat{D}_i$ where $\hat{D}_i = \exp\left[ \sum_\alpha \sqrt{S_{\alpha}} \left(
    \hat b^\dagger_{\alpha,i} - \hat b^{}_{\alpha,i} \right) \right] $ with
$\hat{b}_{\alpha,i}$ the annihilation operator for the $\alpha$th
rovibrational mode of molecule $i$, and $\sqrt{S_{\alpha}}$ the corresponding
Huang-Rhys parameter.  In the weak coupling limit, the matter-light
coupling can be treated perturbatively, leading to incoherent emission
and absorption processes, described in a master equation by terms
\begin{multline}
  \dot \rho = \ldots +
  \sum_{\modelabel,i} 
  \Gamma(-\delta_\modelabel = \epsilon - \omega_\modelabel)
  \mathcal{L}[\hat a^\dagger_\modelabel \sigma_i^-, \rho]
  \\+
  \Gamma(\delta_\modelabel = \omega_\modelabel - \epsilon)
  \mathcal{L}[\hat a^{}_\modelabel \sigma_i^+, \rho].
\end{multline}
Here the function $\Gamma(\delta)$ can be calculated from the Fourier
transform of the two-time correlation function of the operators $\hat D_i(t)$ and
$\hat D_i^\dagger(t^\prime)$.  Further details are given in
Ref.~\cite{Kirton2015}.  Crucially, for thermalized vibrational degrees of
freedom one finds the
Kennard-Stepanov~\cite{Kennard1918,Kennard1926,Stepanov1957} relation
$\Gamma(-\delta) = e^{-\beta \delta} \Gamma(\delta)$.  Alternatively, the
function $\Gamma(\delta)$ can be found experimentally from the observed
fluorescence of the dye.  We will use this approach, and specifically the
spectrum extracted from experimental measurement of Rhodamine 6G~\cite{Nyman:data}, as in Ref.~\cite{keeling16} here.

To consider the polarization state of the light there are two changes we
must make.  We must obviously keep track separately of the different
polarization components of the light, and we must also take account of the
orientation of the dipole moments of the molecules.  We consider a
  limit where the molecules are strongly anisotropic, so there is only a single
  non-degenerate electronic excited state that is relevant, and a single
  associated dipole moment for the ground to excited state transition.
  (For a situation where the molecule is more spherically symmetric,
  would require one to keep track of multiple electronic excited states, and
  the orientation of the dipole moment for transitions to each state
  separately.)  Accounting for polarization leads to a modified
Hamiltonian, of the form
\begin{multline}
  \label{eqn:Hpol}
  \hat{H}=
  \sum_{\modelabel, \sigma={x,y}}
  \omega_\modelabel\hat{a}^\dagger_{\sigma\modelabel} \hat{a}^{}_{\sigma\modelabel} 
  +
  \sum_i {\omDye} \sigma^+_i \sigma^-_i
  \\+
  g\sum_{\modelabel,\sigma,i}\hat{\vec{e}}_\sigma \cdot \hat{\vec{d}}_i \left(
    \hat{a}_{\sigma\modelabel}\hat{\sigma}_i^+
    +
    \hat{a}^\dagger_{\sigma\modelabel}\hat{\sigma}_i^-
  \right).
\end{multline}
In comparison to Eq.~(\ref{eqn:H0}), the crucial extra feature in
Eq.~(\ref{eqn:Hpol}) is
that we take account of the orientation $\vec{d}_i$ of the dipole moment of 
molecule $i$, and how this affects its coupling to light with polarization
$\hat{\vec{e}}_{\sigma=x,y} = \hat{\vec{x}}, \hat{\vec{y}}$.  
As illustrated in Fig.~\ref{fig:cartoon}, the orientation of molecule $i$ can 
be parameterized by the polar angles $\theta_i$, $\phi_i$ describing the
orientation  with respect to the cavity axis,
and yields $\hat{\vec{e}}_{\sigma=x,y} \cdot \hat{\vec{d}}_i = \sin\theta_i
\times (\cos \phi_i, \sin \phi_i)$ respectively.  The intra-molecular coupling
between the electronic state and vibrational state of a given molecule is not
dependent on how the molecule is oriented with respect to any external axis,
and so its treatment remains the same as our earlier model that
ignored polarization.  In addition to these incoherent processes considered
previously, we will also add one extra crucial process: rotational diffusion
of the molecules. This means we will assume the orientation of each molecule varies
randomly.  If this process is fast, the fluorescence of the molecules is
unpolarized even for a polarized pump, as the orientation of molecules when
they emit and when they absorb becomes uncorrelated.

The model we consider neglects any direct interactions between
  different dye molecules.  i.e., there is no F\"orster resonance energy
  transfer process between molecules.  Such an assumption is reasonable
  for current experiments, where the dye concentrations mean the typical
  distance between molecules exceeds $10$nm, but may play a r\^ole in
  experiments with other materials or at higher concentrations.

\subsection{Equations of motion in angle space}
\label{sec:equat-moti-angle}

Rather than considering the full (and computationally intractable) dynamics of
the photon and molecule density matrix, we follow the same approach as used
previously~\cite{Kirton2015,Kirton2013b} and consider a semiclassical analysis.
Neglecting polarization, the state of the system is described by
two types of quantity: the population $ n_\modelabel = \langle \hat{a}^\dagger_\modelabel
\hat{a}^{}_\modelabel \rangle$ of photons in a given mode, $\modelabel$, and the number
$N_\uparrow = \sum_i \langle \sigma^+_i \sigma^-_i \rangle$ of molecules in
the excited state.  When we account for the polarization of light, and
its selective coupling to molecules with dipole moments oriented in a given
direction, both of these become more complicated.  For light we must now
consider the co- and cross-polarization components $ n^{\sigma\sigma^\prime}_\modelabel
= \langle \hat{a}^\dagger_{\sigma \modelabel} \hat{a}^{}_{\sigma^\prime \modelabel} \rangle$,
while for molecules we have to consider not only the total population, but its
angular distribution $N_\uparrow(\theta,\phi) = \sum_i \delta(\theta-\theta_i)
\delta(\phi-\phi_i) \langle \sigma^+_i \sigma^-_i \rangle$, which in the
large $N$ limit can be considered as a continuous function.

\begin{widetext}
Taking into account all the processes described above, the equations of
motion for the populations and coherences of the cavity modes take the
form:
\begin{align}
  \label{eq:photon_xx}
  \frac{\partial}{\partial t} \, n_{\modelabel}^{x x} &= - \kappa \, n_{\modelabel}^{x x} 
  + \dsolid \, \sin^{2}(\theta) \Bigg \{ 
  \Gamma(-\delta_{\modelabel}) \bigg [ \cos^2(\phi) ( n_{\modelabel}^{x x} + 1 ) 
  + \frac{1}{2} \sin(\phi) \cos(\phi) \Big ( n_{\modelabel}^{x y} + \overline{n_{\modelabel}^{x y}} \Big ) \bigg ] N_{\uparrow}(\theta, \phi) 
  \nonumber\\&\qquad\qquad\qquad\qquad\qquad\qquad
  - \Gamma(\delta_{\modelabel}) \bigg [ \cos^2(\phi) \, n_{\modelabel}^{x x} 
  + \frac{1}{2} \sin(\phi) \cos(\phi) \Big ( n_{\modelabel}^{x y} + \overline{n_{\modelabel}^{x y}} \Big ) \bigg ] 
  \Big( N - N_{\uparrow}(\theta, \phi) \Big) \Bigg \}
  \\
  \label{eq:photon_yy}
  \frac{\partial}{\partial t} \, n_{\modelabel}^{y y} &= - \kappa \, n_{\modelabel}^{y y}
  + \dsolid \, \sin^{2}(\theta) \Bigg \{ 
  \Gamma(-\delta_{\modelabel}) \bigg [ \sin^{2}(\phi) ( n_{\modelabel}^{y y} + 1 ) 
  + \frac{1}{2} \sin(\phi) \cos(\phi) \Big ( n_{\modelabel}^{x y} + \overline{n_{\modelabel}^{x y}} \Big ) \bigg ] N_{\uparrow}(\theta, \phi) 
  \nonumber\\&\qquad\qquad\qquad\qquad\qquad\qquad
  - \Gamma(\delta_{\modelabel}) \bigg [ \sin^{2}(\phi) \, n_{\modelabel}^{y y} 
  + \frac{1}{2} \sin(\phi) \cos(\phi) \Big ( n_{\modelabel}^{x y} + \overline{n_{\modelabel}^{x y}} \Big ) \bigg ] 
  \Big( N - N_{\uparrow}(\theta, \phi) \Big) \Bigg \}
  \\
  \label{eq:photon_xy}
  \frac{\partial}{\partial t} \, n_{\modelabel}^{x y} &= - \kappa \, n_{\modelabel}^{x y} 
  + \frac{1}{2} \dsolid \, \sin^{2}(\theta) \Bigg \{ 
  \Gamma(-\delta_{\modelabel}) \bigg [ n_{\modelabel}^{x y} + \sin(\phi) \cos(\phi) \Big ( n_{\modelabel}^{x x} + n_{\modelabel}^{y y} + 2 \Big ) \bigg ] N_{\uparrow}(\theta, \phi) 
  \nonumber\\&\qquad\qquad\qquad\qquad\qquad\qquad
  - \Gamma(\delta_{\modelabel}) \bigg [ n_{\modelabel}^{x y} + \sin(\phi) \cos(\phi) \Big ( n_{\modelabel}^{x x} + n_{\modelabel}^{y y} \Big ) \bigg ] \Big( N - N_{\uparrow}(\theta, \phi) \Big) \Bigg \}
\end{align}
where we have written $\dsolid = \int_{0}^{2 \pi} d \phi \int_{0}^{\pi} d
\theta \, \sin(\theta)$ for the integral over solid angles, and
the bar indicates complex conjugation.   In each
of these equations there are two types of processes: simple cavity loss, and
emission and absorption from and by the dye molecules. The location
of the $+1$ terms can be understood as arising from the
commutator $[\hat a^{}_{\modelabel \sigma}, \hat a^\dagger_{\modelabel \sigma^\prime}]$.

To complete the description, we need also an equation of motion for the
angular distribution of the excited molecules.  Note, we are assuming angular
diffusion of molecules is independent of the electronic state.  This means
that the angular distribution of all molecules, $N_\uparrow(\theta,\phi) +
N_\downarrow(\theta,\phi) = N$ is a constant, independent of angle, hence the
terms $N- N_\uparrow(\theta,\phi)$ appearing above indicate the angular
distribution of ground state molecules.  Thus, we need only track the
evolution of the excited state distribution, $N_\uparrow(\theta,\phi)$ which
obeys:
\begin{align}
\label{eq:ang-dist}
  \frac{\partial}{\partial t} \, N_{\uparrow}(\theta,\phi) &= 
  - \Gamma_{\downarrow} \, N_{\uparrow}(\theta,\phi)
  +  
  \Gamma_{\uparrow} \sin^2(\theta) 
  \Big( \cos^{2}(\chi) \cos^{2}(\phi) + \sin^{2}(\chi) \sin^{2}(\phi)  \Big)
  \Big( N - N_{\uparrow}(\theta,\phi) \Big) \nonumber\\
  &
  + \sum_{\modelabel=0}^{\infty} g_{\modelabel} \,  
\sin^2(\theta) \bigg \{
  \Gamma(\delta_{\modelabel}) \bigg[ \cos^{2}(\phi) n_{\modelabel}^{x x} + \sin^{2}(\phi) n_{\modelabel}^{y y} + \sin(\phi) \cos(\phi) \big ( n_{\modelabel}^{x y} + \overline{n_{\modelabel}^{x y}} \big ) \bigg ] \Big ( N - N_{\uparrow}(\theta,\phi) \Big ) 
  \nonumber\\
  &\qquad\qquad
  - \Gamma(-\delta_{\modelabel}) \bigg[ \cos^{2}(\phi) \big ( n_{\modelabel}^{x x} + 1 \big ) + \sin^{2}(\phi) \big ( n_{\modelabel}^{y y} + 1 \big ) + \sin(\phi) \cos(\phi) \big ( n_{\modelabel}^{x y} + \overline{n_{\modelabel}^{x y}} \big ) \bigg ] N_{\uparrow}(\theta,\phi) \bigg \}  \nonumber\\
  &
  + D \bigg [ \frac{1}{\sin(\theta)} \frac{\partial}{\partial \theta} \Big ( \sin(\theta) \frac{\partial}{\partial \theta} \Big ) + \frac{1}{\sin^{2}(\phi)} \frac{\partial^{2}}{\partial \phi^{2}} \bigg ] N_{\uparrow}(\theta,\phi).
\end{align}
where $g_\modelabel$ accounts for the degeneracy of a given photon mode energy $\modelabel$;
in the following we always consider a two-dimensional harmonic
oscillator mirror profile, so that $g_\modelabel = \modelabel+1$.
\end{widetext}

In Eq.~(\ref{eq:ang-dist}), the first line represents the external
(non-cavity-mediated) loss, and external pumping process acting on the
molecules.  We consider a pump which has total intensity $\Gamma_\uparrow$,
and a polarization which varies from fully polarized in the $x$ direction
(for $\chi=0$), through fully unpolarized ($\chi=\pi/4$), to fully
polarized in the $y$ direction ($\chi=\pi/2$).  In terms of the Bloch
sphere of polarization, this means that as $\chi$ is varied, one follows
an axis through the center of the sphere. Since pumping of a given
molecule depends on the overlap of its dipole moment with the polarization
of this pump laser, this leads to the form of the pump term seen.  In
contrast, the non-cavity-mediated decay (e.g.\ via non-radiative processes)
is assumed independent of the orientation of the molecules.  The second and
third line of Eq.~(\ref{eq:ang-dist}) are the counterpart of the photon
rate equations, describing how emission and absorption of cavity light
affects the angular distribution of excited molecules.  The final line is
the rotational diffusion process which we write as a Laplacian with a
rate constant $D$; these rotational rate constants for a variety of dye
molecules in various solvents have been measured in~\citet{VonJena79}.

Given the form of external pumping we have considered in the equations above,
we may note that at late times two interconnected simplifications occur: 
$n_{\modelabel}^{xy} \to 0$, and
$N_\uparrow(\theta,\phi)$ becomes an even function of $\phi$.  If we consider
the second condition first, we note that in Eq.~(\ref{eq:ang-dist}), that if
$N_\uparrow(\theta,\phi)$ is an even function of $\phi$, then the only odd
terms on the right hand side come from $n^{xy}_\modelabel$, with its $\sin(\phi)
\cos(\phi) = \sin(2\phi)/2$ dependence.  If we then consider
Eq.~(\ref{eq:photon_xy}), we see that the source term for $n_\modelabel^{xy}$
depends precisely on the integral $\dsolid \sin^2(\theta) \sin(2\phi)
N_\uparrow(\theta,\phi)$.  Thus, if $N_\uparrow(\theta,\phi)$ is an even
function, this source term vanishes.  Since diffusion of the molecules causes
decay of all angular dependence of $N_\uparrow(\theta,\phi)$, and photon loss
causes decay of $n_\modelabel^{xy}$, we can see that any initial odd harmonics
or population of $n_\modelabel^{xy}$ will be lost.   

The vanishing steady state value of $n_\modelabel^{xy}$ is a consequence of
our choice of pump polarization: because our pump contains an incoherent
mixture of $x$ and $y$ polarized light, it cannot break the phase symmetry for
the complex $n_\modelabel^{xy}$.  If we had chosen alternate axes for the
linear polarization components of the pump, then the angular distribution of
excited molecules would have contained odd components, leading to a non-zero
values of $n_\modelabel^{xy}$.  However, such a situation could be reduced to
the one we consider by a linear rotation of polarization axes.

One notable consequence of the equations of motion, which we will discuss
below, is that even in the absence of diffusion, perfect linear polarization
of the pump does not lead to perfect polarization of the cavity light.
Microscopically this is because molecules oriented at $\phi=\pi/4$ can couple
with $\sin^2 \phi = \cos^2\phi=1/2$ to both $x$ and $y$ polarized light.
Naively one might have expected that without diffusion, the orthogonal $x$ and
$y$ polarizations would decouple. For the molecules to preserve this orthogonality 
requires inter-molecular
coherence.  However, due to frequent collisions between the dye molecules and the solvent this coherence vanishes (our equations are written using this assumption) and so the destructive interference does not occur.


\subsection{Equations of motion for angular harmonics}
\label{sec:equat-moti-angul}

Making use of the above simplification, we have two sets of equations for the
photon populations and a partial differential equation for the angle
distribution.  To solve this numerically, it is helpful to rewrite this
distribution in terms of standard spherical harmonics.  Specifically we write:
\begin{align*}
  N_\uparrow(\theta,\phi) &= 
  \sum_{l,m} N_{l,m}
  Y_{l,m}(\theta,\phi),
  \\  
  Y_{l,m}(\theta,\phi)
  &=
  \sqrt{\frac{2l+1}{4\pi} \frac{(l-m)!}{(l+m)!}}
  P_{l,m}(\cos\theta) e^{i m \phi}
\end{align*}
where $P_{l,m}(x)$ are the associated Legendre polynomials.  

\begin{widetext}
  
Written
in this way, the photon population equations take a relatively
simple form:
\begin{multline}
  \label{eq:photon_sh}
  \frac{\partial}{\partial t} \, n_{\modelabel}^{x x, y y} 
  = - \kappa \, n_{\modelabel}^{x x,y y} + \frac{4 \pi}{3} \Bigg[ 
  \Gamma(- \delta_\modelabel) \, \bigg \{  \big(n_{\modelabel}^{x x,y y} + 1 \big) \Big [ N_{0,0} - \frac{N_{2,0}}{\sqrt{5}}   \pm \sqrt{\frac{3}{10}} \, \big ( N_{2,2} + N_{2,-2} \big ) \Big ]  \bigg \} 
  \\
  - \Gamma(\delta_\modelabel) \bigg \{ n_{\modelabel}^{x x,y y} \, \Big [ N - N_{0,0} + \frac{N_{2,0}}{\sqrt{5}}  \mp \sqrt{\frac{3}{10}} \, \big ( N_{2,2} + N_{2,-2} \big ) \Big ]  \bigg \} \Bigg].
\end{multline}
Here, the only difference between the $xx$ and $yy$ populations is the sign 
of the populations of the $m=\pm2$ harmonics.  One may note that only the
even-parity harmonics with $l \in \{0, 2\}$ actually
couple to the photon distribution.  However, as we will see next, there
are (photon induced) couplings between different harmonic
components of the molecular distribution.

The equation of motion for $N_{l,m}$ takes a more complicated form, so we
introduce various auxiliary quantities $\zeta^\pm_{l,m},\mu^\pm_{l,m}$,
defined below. In terms of these we have:
\begin{multline}
\label{eq:molecule_sh}
  \frac{\partial}{\partial t} \, N_{l,m} = 
  - \, \Gamma_{\downarrow} \, N_{l,m} 
  +  \Gamma_{\uparrow} \, 
  \left( \cos^2(\chi) \mu_{l, m}^{+} + \sin^2(\chi) \mu_{l,m}^- \right) \\
  + \sum_{\modelabel=0}^{\infty} g_{\modelabel} \bigg[ 
  \Gamma(\delta_\modelabel) \, 
  \left( n_\modelabel^{xx} \mu_{l, m}^{+} + n_\modelabel^{yy} \mu_{l, m}^{-} \right) - 
  \Gamma(-\delta_\modelabel)  \, 
  \left( \left( n_{\modelabel}^{xx} + 1 \right) \zeta_{l,m}^{+} 
       + \left( n_{\modelabel}^{yy} + 1 \right) \zeta_{l,m}^{-} \right) \bigg] 
  - D \, l  \left( l + 1 \right) \, N_{l,m}.
\end{multline}
Note here that the label $\modelabel$ denotes cavity modes, while
$m$ denotes azimuthal harmonics of the molecular distribution.
The auxiliary quantities are defined by:
\begin{equation}
  \label{eq:mu}
  \mu_{l, m}^{\pm} = \frac{N}{3} \left[ 
    \delta_{0, l} \delta_{0, m} - \frac{1}{\sqrt{5}} \, \delta_{2, l} \delta_{0, m} 
    \pm \sqrt{\frac{3}{10}} \, 
    \left(
      \delta_{2,l} \delta_{2, m} + \delta_{2, l} \delta_{-2, m} 
    \right)
  \right]
  - \zeta_{l, m}^{\pm},
\end{equation}
and
\begin{multline}
  \label{eq:zeta}
  \zeta_{l, m}^{\pm} = \frac{1}{2} \Bigg [
   \, \bigg ( 1 - \frac{(l-m+1)(l+m+1)}{(2l+1)(2l+3)} - \frac{(l+m)(l-m)}{(2l+1)(2l-1)} \bigg ) \, N_{l, m} 
  -  \sqrt{\frac{(l-m)(l-m-1)(l+m)(l+m-1)}{(2l-3)(2l-1)^2(2l+1)}} \, N_{l-2, m} \\
  -  \sqrt{\frac{(l-m+2)(l-m+1)(l+m+2)(l+m+1)}{(2l+1)(2l+3)^2(2l+5)}} \, N_{l+2, m} 
  \\
  \pm \pi \,  \sqrt{\frac{(2 l + 1) \, (l - m)!}{4 \pi \, (l + m)!}} \, \sum_{l^{\prime} = 0}^{\infty} 
  \sum_{m^\prime = m-2, m+2}
  \sqrt{\frac{(2 l^{\prime} + 1) \, (l^{\prime} - m^\prime)!}{4 \pi \, (l^{\prime} + m^\prime)!}} \int_{-1}^{1} dx \, \Big( 1-x^{2} \Big) \, 
  P_{l, m}(x) \, P_{l^{\prime}, m^\prime}(x) \ N_{l^{\prime}, m^\prime} 
\Bigg ].
\end{multline}
Equations~(\ref{eq:photon_sh})--(\ref{eq:zeta}) define the problem that we
simulate numerically.  From this we can then extract the full spectrum of the
photons $n^{\sigma\sigma}_{\modelabel}$, and thus quantities such as the total intensity of
$x$ and $y$ polarized light, or the total degree of linear polarization. We truncate the equations at a fixed value of $l$ and, to ensure convergence, we check that all results are insensitive to increasing this cutoff.  
\end{widetext}

\subsection{Parameters}
\label{sec:parameters}

In the following, we show numerical results of these equations, and so must
specify parameters.  For reference we discuss here the typical parameter
values used.  Table~\ref{tab:parameters} provides the values of these
parameters.  Note that $\Gamma_\uparrow$ is not specified, as we typically
scan the pumping strength in order to cross threshold, and we report the
ratio $\Gamma_\uparrow/\Gamma_\downarrow$.  The polarization angle of the pump, $\chi$,
is always given in all figures (or ``full polarization'', i.e.\ $\chi=0$ is
stated).

\begin{table}[htpb]
  \centering
  \begin{tabular}{|c|p{0.25\textwidth}|l|}
    \hline
    \bf Parameter & \bf Meaning & \bf Value \\
    \hline
    $\kappa$ & Cavity mode decay rate &\SI{0.5}{\GHz} \\
    $\Gamma_\downarrow$ & (Non-cavity) Decay rate of excited electronic 
    state & \SI{0.25}{\GHz} \\
     $D$ & Rotational diffusion rate of molecules & \SI{0.333}{\GHz} \\
     $N$ & Number of molecules & $10^8$  \\   
    \hline
  \end{tabular}
  \caption{Standard parameter values used in figures below unless otherwise specified.}
  \label{tab:parameters}
\end{table}

As well as these simple parameters, a crucial parameter choice is the set
of values $\Gamma(\delta_\modelabel)$ for the different photon modes.  We take the
same functional form of $\Gamma(\delta)$ as used in Ref.~\cite{keeling16},
corresponding to the shape of the experimentally measured spectrum of
Rhodamine 6G~\cite{Nyman:data}, with a peak height around \SI{10}{\kHz}, extracted in
Ref.~\cite{keeling16} by matching the pump-spot-size dependence of photon
cloud below threshold.  We
then sample at a set of frequencies $\delta_\modelabel = \delta_0 + \modelabel \epsilon$
where the cavity cutoff frequency is $\delta_0=$ \SI{3300}{\THz} and mode spacing
(set by the mirror curvature) of $\epsilon =$ \SI{2.7}{\THz}.  These parameters
mean we are in a regime where the system thermalizes well (i.e.\ this 
 is  significant emission and absorption right down to the 
cavity cutoff).  It is also important to note that comparing this form
of $\Gamma(\delta)$ to the values $\Gamma_\downarrow$ means that at low
powers (without stimulated emission), non-cavity loss processes are
dominant, leading to a sharp threshold (see Ref.~\cite{keeling16} for
further details).

\section{Results}
\label{sec:results}

\subsection{Characteristic steady state mode occupations}
\label{sec:char-steady-state}

\begin{figure}
  \centering
  \includegraphics[width=3.2in]{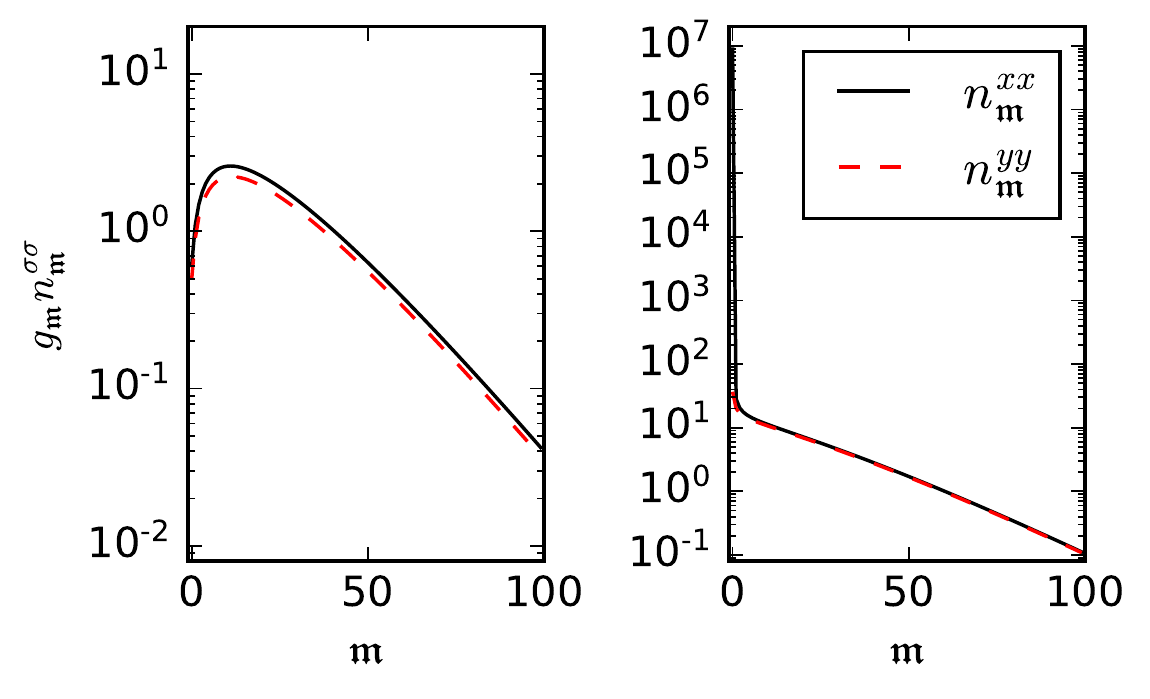}
  \caption{Occupation of photon states for a fully polarized pump.  Left:
    Below threshold, Right: above threshold.  The two lines show the two
    different polarization states $n_\modelabel^{xx}, n_\modelabel^{yy}$.  See
    table~\ref{tab:parameters} for parameter values.}
  \label{fig:occupations}
\end{figure}

Before discussing the dependence of the polarization on external parameters we show the steady state of our equations of
motion, plotted in Fig.~\ref{fig:occupations} for two pump powers, one below
threshold and the other above threshold.  These are plotted for a fully
polarized pump, $\chi=0$,  with a non-zero rotational diffusion constant
$D$ and  other parameters  as given in Table~\ref{tab:parameters}.

In both the below- and above-threshold cases, the distribution closely matches
the Bose-Einstein distribution, but above threshold, it is in the regime where
a Bose condensed fraction of photons arises.  As is clear from the figure,
there is a considerable change of the degree of polarization between the
below-threshold and above-threshold behavior.  This has a simple theoretical
explanation: below threshold, the state is almost unpolarized, as the
rotational diffusion randomizes the orientation of the molecule between
absorption and emission.  i.e., the timescale for molecules to rotate is much
shorter than the timescale for fluorescence. Thus, despite the polarized pump,
the subsequent fluorescence of the molecules produces a nearly unpolarized
source of photons in the cavity.  Above threshold, the macroscopic population
of photons in the low energy mode leads to stimulated emission of photons into
that mode.  That means the rate of emission of photons increases, by a factor
depending on the occupation of the condensate mode.  This increased emission
rate means the fluorescence becomes faster than the rotation and so polarization
is better preserved.  In the following we will systematically explore
the dependence of this process on various parameters.

\subsection{Polarization degree across condensation threshold}
\label{sec:polar-degr-across}

\begin{figure}
  \centering
  \includegraphics[width=3.2in]{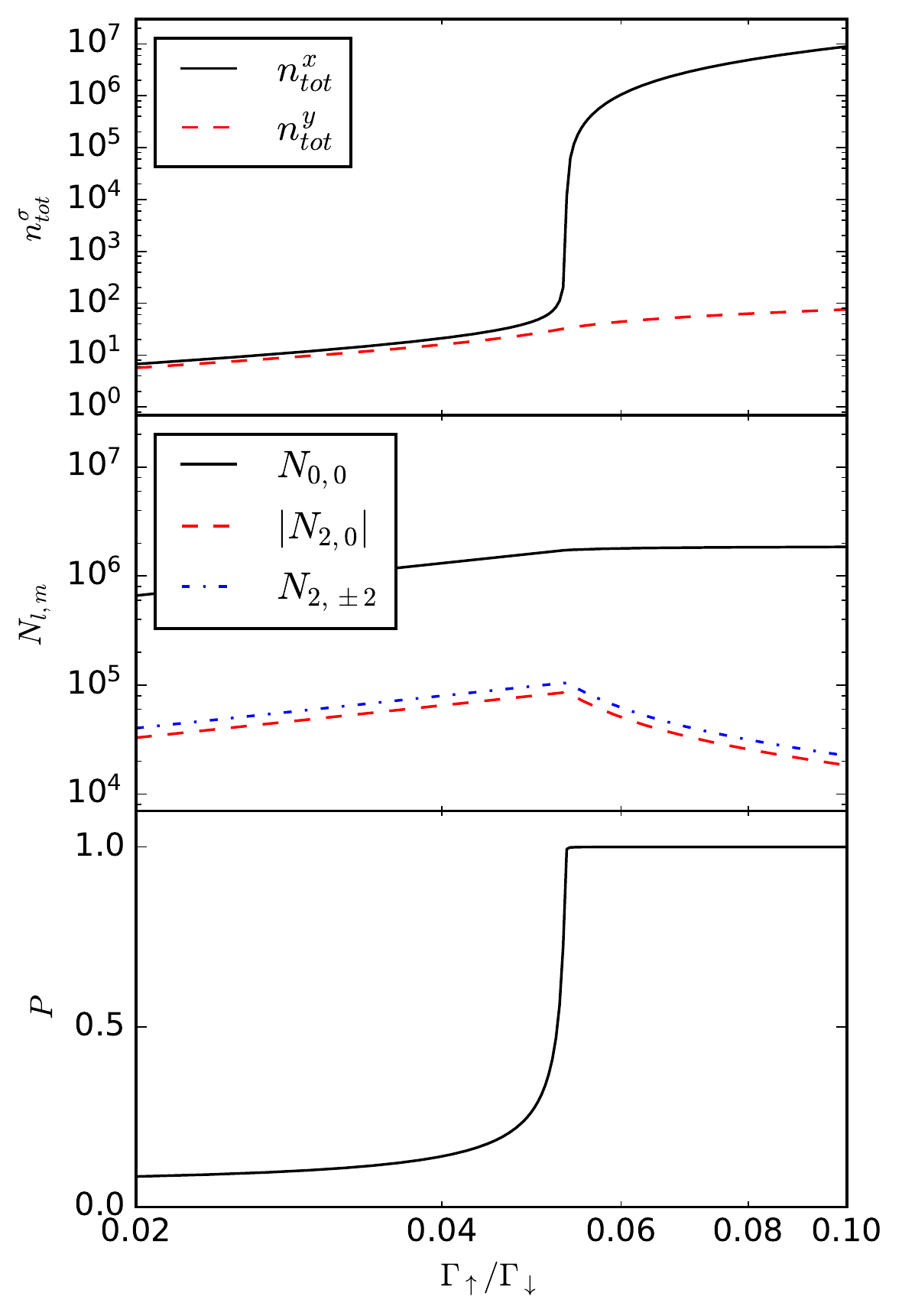}
  \caption{Total polarization as a function of dimensionless pump strength $\Gamma_\uparrow/\Gamma_\downarrow$ for a fully $x$ polarized pump. The top panel
    shows total intensity $\ntot^{\sigma} = \sum_\modelabel n^{\sigma \sigma}_\modelabel$ as a
    function of pump strength for $\sigma=x,y$. The middle panel shows the population of molecules in each of the lowest order spherical harmonics. Bottom panel shows the
    corresponding value of $P = (\ntot^x - \ntot^y)/(\ntot^x + \ntot^y)$.  We
    see that at the condensation threshold, the degree of polarization
    increases significantly.}
    \label{fig:pol_vs_pump}
\end{figure}

In order to investigate how the polarization changes from below to above
threshold, and to orient further discussion, Fig.~\ref{fig:pol_vs_pump}
summarizes the degree of polarization by first plotting the total light
intensity $\ntot^{\sigma=x,y} = \sum_\modelabel n^{\sigma \sigma}_\modelabel$ vs pump
strength, and then the polarization degree $P = (\ntot^x -
\ntot^y)/(\ntot^x + \ntot^y)$, which varies between $P=1$ for a fully $x$
polarized condensate, to $P=0$ for a fully unpolarized state, to $P=-1$ for
fully $y$ polarized.  As anticipated above, for a fully polarized pump, the degree of polarization
increases significantly at the same point that the total photon population
changes from increasing linearly with pump to superlinearly, i.e.\ at the
point where macroscopic occupation of a single mode, and thus stimulated
emission sets in. 

The middle panel of Fig.~\ref{fig:pol_vs_pump} shows the corresponding
behavior of three of the angular moments of the molecular distribution,
$N_{0,0}$, $N_{2,0}$ and $N_{2,2} = N_{2,-2}$.  The lowest order component
corresponds to an overall scale of the molecular excitation, and as one may
expect, it increases with pumping below threshold, and then saturates at
threshold.  More notably, for the components with $l=2$, these decrease
above threshold.  When both polarization components are above threshold,
the chemical potential of light, $\mu$ must be locked at the frequency of
the lowest cavity mode, i.e.\ $\mu=\delta_0$.  If the molecular distribution
were to come into equilibrium with this, then one would require an
equilibrated  distribution $N_\uparrow(\theta,\phi) = [ e^{-\beta
  \mu} + 1]^{-1}$ independent of angle, 
which would imply $N_{l,m} \to 0$ for all $l, m > 0$.
This behavior is the polarization equivalent of gain saturation leading to
a spatially flat excitation profile in an inhomogeneous pump
spot~\cite{keeling16}.

\subsection{Dependence on cavity loss rate}
\label{sec:depend-cavity-loss}

\begin{figure}
  \centering
  \includegraphics[width=3.2in]{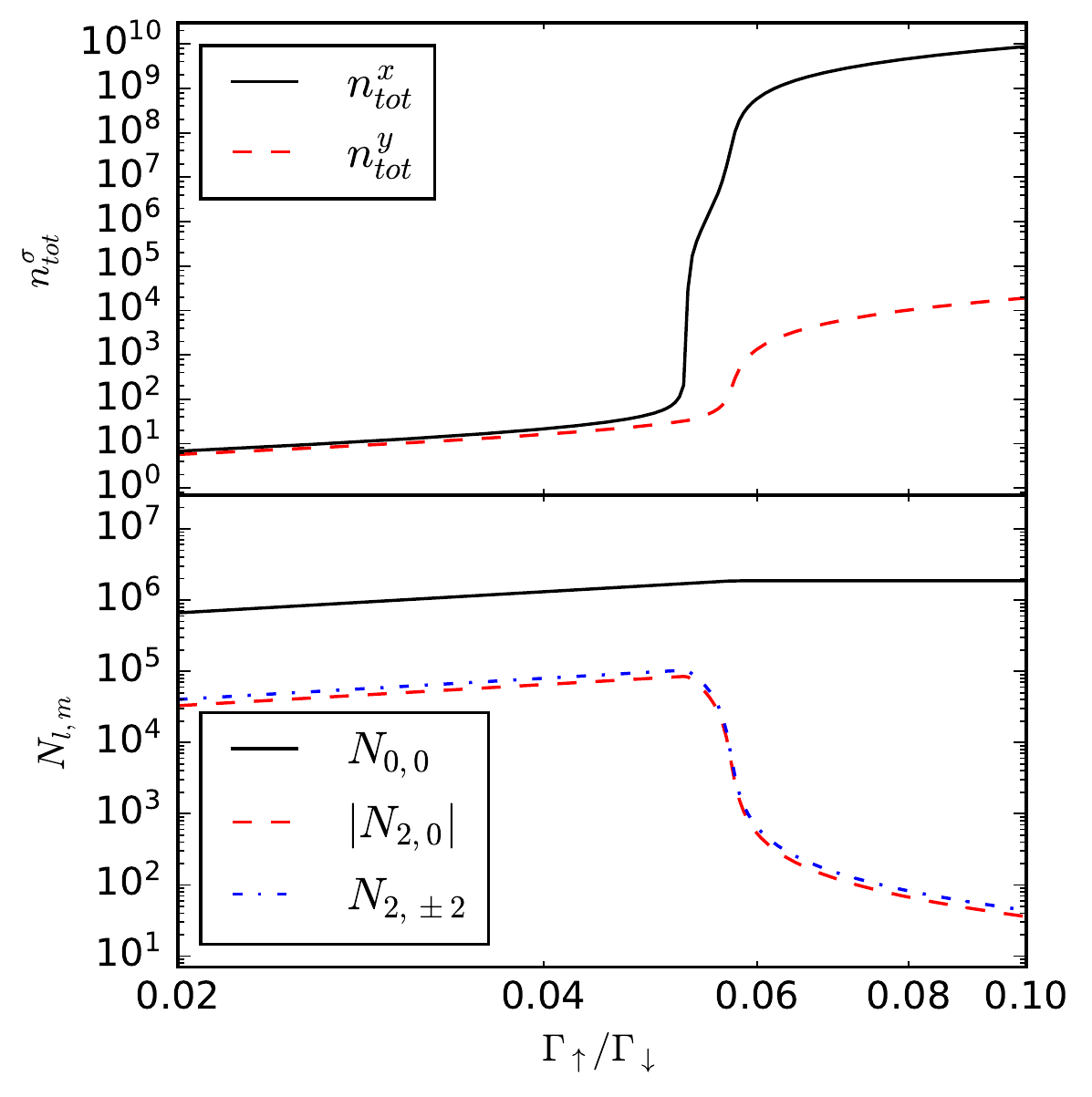}
  \caption{Total intensity of each polarization vs dimensionless pump power as in
    Fig.~\ref{fig:pol_vs_pump}, but for a smaller loss rate,
    $\kappa=$ \SI{0.5}{\MHz}.  All other parameters as in
    Fig.~\ref{fig:pol_vs_pump}.  With a smaller value of $\kappa$, both
    polarizations go above threshold.}
    \label{fig:pol_vs_pump_small_kappa}
\end{figure}

In Figure~\ref{fig:pol_vs_pump}, over the range of pumping shown, only the
majority polarization component acquires a macroscopic population.
However, this is not always the case.  For comparison,
Fig.~\ref{fig:pol_vs_pump_small_kappa} shows the same quantities as
Fig.~\ref{fig:pol_vs_pump} but for a smaller cavity loss rate $\kappa$.  In
this case, one sees that both components reach threshold (both components
show a nonlinear increase of population at a critical pumping strength).
In contrast, if we increase the cavity loss rate, the population of the
minority component is  reduced.
Multiple modes reaching threshold is not inherently surprising: multimode
behavior has been predicted~\cite{keeling16} and
observed~\cite{marelic16:njp} in the dye-cavity system due to spatial hole
burning.  In that case, spatial hole burning allows non-degenerate
 modes with different
transverse profiles to reach threshold at a higher pump power than the
first lasing mode.  Since the two different polarization states are
degenerate, the bare emission rates into these modes are equal, and so both
modes can in principle reach threshold at the same power.

The above discussion might suggest that in a perfect cavity, i.e.\ in the
limit of vanishing photon loss, polarizations would vanish.  However,
finite polarization can remain in this limit, despite the fact that the two
polarizations are degenerate in energy.  To see this, we may consider
Eq.~(\ref{eq:photon_sh}), from which we see that 
the steady state photon distribution must obey:
\begin{align}
  \frac{n_\modelabel^{xx,yy} + 1}{n_\modelabel^{xx,yy}}
  = 
  \frac{\Gamma(\delta_\modelabel)}{\Gamma(-\delta_\modelabel)}
  \frac{N-N^{xx,yy}}{N^{xx,yy}}
\end{align}
where we have denoted $N^{xx,yy} = N_{0,0} - {N_{2,0}}/{\sqrt{5}} \pm
\sqrt{\frac{3}{10}} ( N_{2,2} + N_{2,-2} )$.  Using the Kennard-Stepanov
relation between $\Gamma(\pm \delta_\modelabel)$, this expression clearly
leads to a Bose-Einstein distribution $n_{\modelabel}^{xx,yy} = \left[
  \zeta_{xx,yy}^{-1} e^{\beta \delta_\modelabel} - 1\right]^{-1}$ where the fugacity
$\zeta_{xx,yy}$ is given by $\zeta_{xx,yy} = N^{xx,yy}/ (N-N^{xx,yy})$.  In
general, $N^{xx} \neq N^{yy}$ (as long as $N_{2,\pm 2}$ is strictly non-zero), 
so these fugacities differ.  Since the
fugacity of the majority component must be close to $\zeta=e^{\beta
  \delta_0}$,  a very small difference in $ N^{xx}$, $N^{yy}$ is sufficient
to sustain a large difference in photon population.  Indeed, as seen in the
central panel of Fig.~\ref{fig:pol_vs_pump} the value of $N_{2,\pm 2}$
actually approaches zero above threshold, but the residual non-zero 
value leads to a finite polarization of the light.

\subsection{Dependence on input polarization}
\label{sec:depend-input-polar}

\begin{figure}
  \centering
  \includegraphics[width=3.2in]{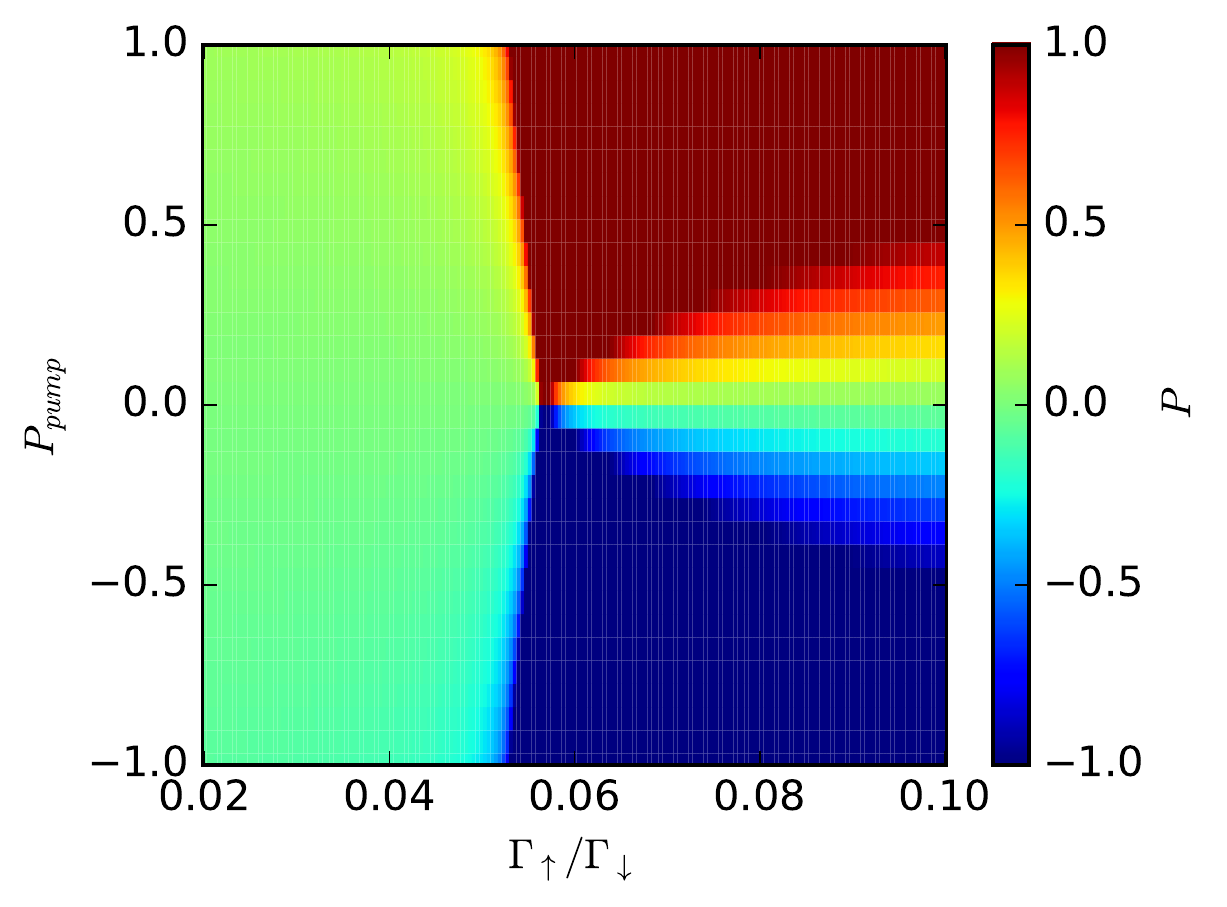}
  \caption{Colormap of polarization degree, $P$ as defined in Fig.~\ref{fig:pol_vs_pump}, as a function of dimensionless pump power (horizontal axis) and polarization degree of pump (vertical axis).}
  \label{fig:colormap-vs-pol}
\end{figure}

Having identified a single parameter $P$ that defines the polarization
state, we use this in Fig.~\ref{fig:colormap-vs-pol} to show how the
evolution of polarization degree with pump power varies according to pump
polarization.  Below threshold, as noted above, the state is very weakly
polarized, and only weakly dependent on the polarization degree of the
light.  Just above threshold, the majority photon polarization grows faster
than the minority, and so there is almost complete polarization in this
limit, with the output polarization switching between $+1$ and $-1$
depending on the sign of the input polarization.  The threshold power
depends on the polarization degree.  This is expected, as a high
polarization degree means that more of the input power can go into feeding
the majority polarization component, and so that component reaches
threshold sooner. Further above threshold (i.e.\ at the
largest values of $\Gamma_\uparrow$ shown in
Fig.~\ref{fig:colormap-vs-pol}), the minority photon polarization also
becomes large, so the polarization degree reduces, leading to a more
gradual dependence on the input polarization. This is particularly noticeable
for smaller input polarization degree.

\subsection{Effect of diffusion constant}
\label{sec:effect-diff-const}

\begin{figure}
  \centering
  \includegraphics[width=3.2in]{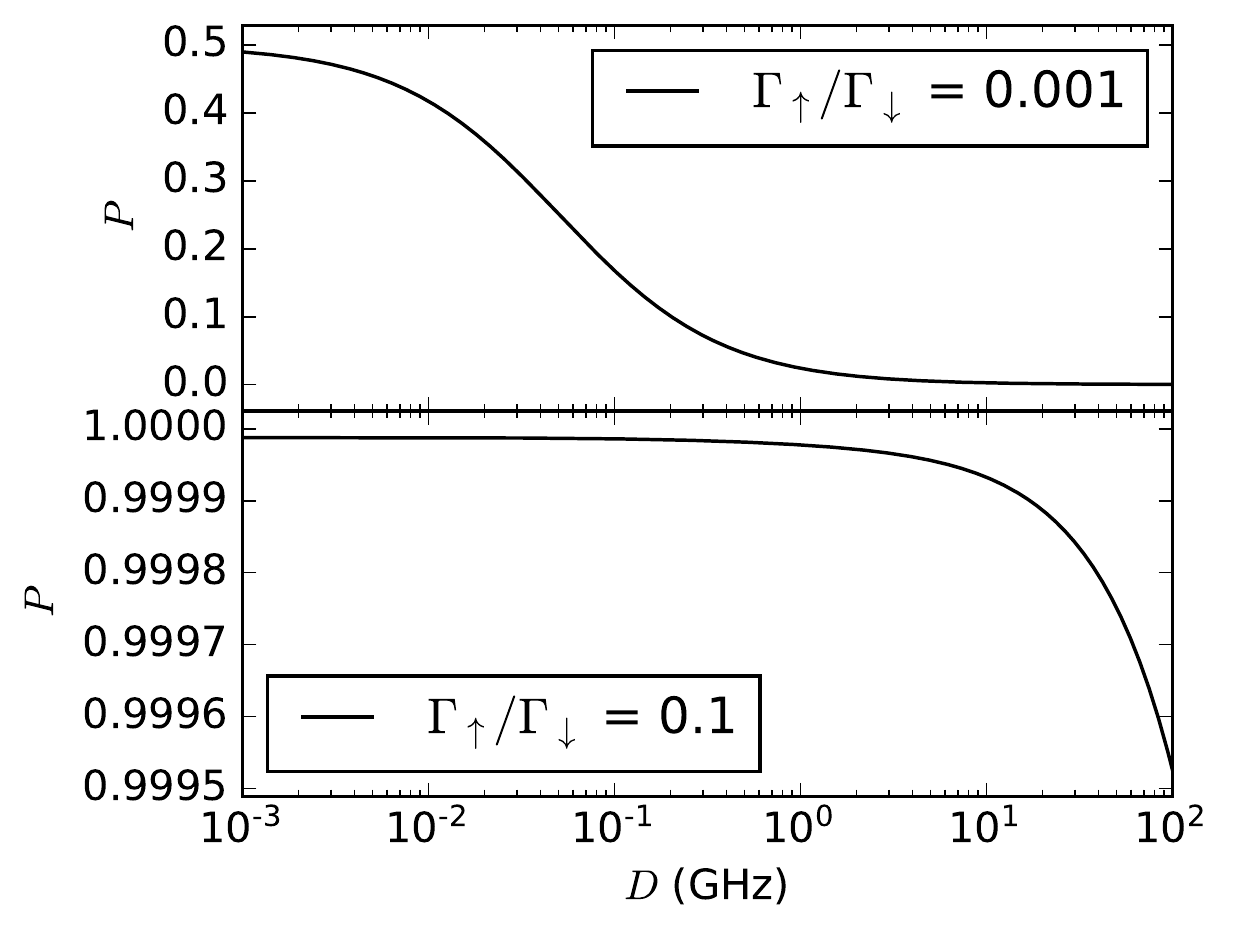}
  \caption{Polarization degree, $P$, as defined in Fig.~\ref{fig:pol_vs_pump} as a
    function of angular diffusion constant of molecules, $D$ for a pump
    power below threshold (top) and above threshold (bottom).}
  \label{fig:colormap-vs-diffusions}
\end{figure}

As noted above, the difference in polarization degree from below to above
threshold originates from the competition between the timescale for
diffusion, and the timescale for (stimulated) fluorescence from the dye
molecules.  Figure~\ref{fig:colormap-vs-diffusions} shows the effect of the
diffusion constant on this.  The behavior is shown both below and above
threshold, considering a fully $x$ polarized pump in both cases.

Well below threshold, there is a gradual increase of the polarization
degree as the diffusion constant reduces.  When the diffusion constant is
zero, there is no molecular rotation; despite this, the system does not
become fully polarized, but reaches a value $P=0.5$.  This is because, as
discussed above, most molecules can couple both to $x$ and $y$ polarized
light.  This means that a purely $x$ polarized pump excites molecules that
can subsequently emit into both $x$ and $y$ polarized modes.

We can understand the limiting value $P=0.5$ by considering the behavior of
Eq.~(\ref{eq:ang-dist}) for weak pumping and zero diffusion.  In the weak
pumping limit, the cavity population is small so all cavity mediated terms
can be neglected.  Considering a fully polarized pump ($\chi=0$), the
steady state of this equation becomes:
\begin{displaymath}
   \Gamma_{\downarrow} \, N_{\uparrow}(\theta,\phi)
  =
  \Gamma_{\uparrow} \sin^2(\theta) \cos^{2}(\phi)
  \left[ N - N_{\uparrow}(\theta,\phi) \right].
\end{displaymath}
Then if $\Gamma_{\uparrow} \ll \Gamma_{\downarrow}$, this equation
implies that $N_{\uparrow}(\theta,\phi) \propto \sin^2(\theta)
\cos^2(\phi)$.  The polarization degree of light then follows from
inserting this form into the integrals in
Eq.~(\ref{eq:photon_xx})-(\ref{eq:photon_yy}).  This gives the ratio of
$n^{xx} : n^{yy}$ in proportion to the overlaps $\int d \phi \cos^2 \phi
\cos^2 \phi : \int d \phi \cos^2 \phi \sin^2 \phi$, which produces a ratio
$3:1$, giving $P=0.5$.  From this derivation, it is clear that this result
only holds in the limit of very weak pumping.  For stronger pumping, the
cavity modes become populated.  This means that there can be stimulated
emission (even below threshold), which favors the majority component,
increasing $P$.  Indeed, at large enough pumping, condensation occurs
leading to $P \to 1$, as seen in the bottom panel of Fig.~\ref{fig:colormap-vs-diffusions}.

It is worth discussing why it is physically the case that molecules can
couple the two orthogonal polarization modes.  This feature (clearly
present in the equations) originates from the fact that we assume no
intra-molecular coherence exists, only incoherent absorption and emission
by molecules.  If the molecules were allowed to retain coherence, emission
by molecules at different angles could lead to destructive interference.
In such a case, an $x$ polarized pump would lead to a pattern of coherence
such that $y$ polarized emission canceled, while $x$ polarized emission
was reinforced.  The rapid dephasing of the molecules in solution prevents
the destructive interference, and so leads instead to the limiting value
$P=0.5$.

\section{Conclusions}
\label{sec:conclusions}

In conclusion, we have developed a model to describe the polarization
states of a BEC of photons in a dye-filled cavity.  This model extends our
previous work~\cite{Kirton2013b,Kirton2015}, accounting for the
polarization states of light, and the effects of angular diffusion of the
dye on the polarization state.

We find distinct behavior above and below threshold.  Below threshold,
photon emission is slow, and so rotational diffusion of the molecules
washes out the pump polarization.  Above threshold, fast stimulated emission
leads to a greater dependence on pump polarization state.  We assume
coherence of the molecules is rapidly lost due to collisions between dye
and solvent molecules.  As a result, a fully polarized pump can always
produce light with the opposite polarization, even without diffusion.  This
is because most molecules can (incoherently) absorb and emit both
polarizations of light, and without coherence, no cancellations can occur.
Thus, for a fully polarized pump far below threshold, the output
polarization varies between $P=0$ (when diffusion is very fast) and $P=0.5$
when diffusion vanishes, and molecules fail to rotate.  In contrast, above
threshold, stimulated emission allows the majority polarization to
dominate, leading to nearly complete polarization, $P \to 1$.

As the photon BEC is a driven dissipative system, it is not surprising that
the pump polarization can have an effect on the output states.  However, it
is notable that our model predicts this dependence survives even in the limit of
vanishing cavity loss, $\kappa\to 0$.  This means there is an absence of
thermalization of polarization. This is in contrast to thermalization between
different spatial modes of the cavity --- here, the $\kappa\to 0$ limit of
our model is known to lead to a perfectly thermal
distribution~\cite{Kirton2015}.  The difference between the thermalization
of spatial modes, and absence of thermalization for polarization modes can
be traced back to the the nature of how the molecules act as a reservoir.
For large pumping spots~\cite{keeling16}, all spatial modes couple to
exactly the same set of molecules as a reservoir.  In contrast, the two
polarizations of light couple differently to molecules at different angles,
meaning that full equilibration need not occur.

The results we present here provide not only a way to model the
polarization dynamics of a photon condensate, but also provide a clear
understanding of why the model behaves as it does.  Our work was
focused on the current experiments where the medium in the cavity is not significantly
birefringent.  An interesting extension of
our work would be to consider birefringent materials, where energetics
could favor one polarization state, and may then compete with that
favored by pumping. The research data supporting this publication can be found
at doi: \href{https://doi.org/10.17630/e9a4fda9-3f3c-4f23-a6fc-227de27c9851}{10.17630/e9a4fda9-3f3c-4f23-a6fc-227de27c9851}.

\begin{acknowledgments}
  We acknowledge helpful discussions with R.\ Nyman and K.\ E.\ Ballantine.
  RIM acknowledges support from the ``Laidlaw Research Internship'' scheme at
  the University of St Andrews.  PK acknowledges support from EPSRC
  (EP/M010910/1).   JK acknowledges support from EPSRC programs ``TOPNES''
  (EP/I031014/1) and ``Hybrid polaritonics'' (EP/M025330/1).
\end{acknowledgments}

\end{document}